\documentclass[11pt,preprint]{aastex}
\usepackage{amsmath}

\usepackage{epsfig}

\shorttitle{Fine-Structure Constant}
\shortauthors{Sumner}

\begin{document}

\title{On the variation of the fine-structure constant in Friedmann Universes}

\author{William Q. Sumner\altaffilmark{1}}

\affil{Computer Science Department, Central Washington University, Ellensburg, WA 98926}

\altaffiltext{1}{sumner@cwu.edu}

\begin{abstract}

The fine-structure constant $\alpha$ does not vary as Friedmann Universes evolve, a conclusion based on assessments of quantum mechanics and electrodynamics.  $\alpha$ is the dimensionless number $\alpha \equiv e^2/{4\pi\varepsilon_o\hbar c}\approx1/137$, where $e$ is the charge of the electron, $\varepsilon_o$ is vacuum permittivity, $c$ is the speed of light, and $\hbar$ is Planck's constant divided by $2\pi$.   This inquiry was motivated by Schr\"odinger's (1939) prediction that all quantum wave functions coevolve with Friedmann geometry and a similar prediction by Sumner (1994) for $\varepsilon_o$.  The functional form of variations in quantum wave functions found by Schr\"odinger is enough to show that $\alpha$  does not vary.  Electrodynamics also predicts that $\alpha$  does not vary.  Evolutionary changes in $c$ exactly cancel those in $\varepsilon_o$ and other factors in $\alpha$ do not change.  Since $\alpha$ appears in all first-order perturbation formulas for atomic energy levels, comparisons of the atomic spectra of distant atoms with those in laboratories provide an experimental measure of this prediction.  Most  experiments find changes in $\alpha$ that are either statistically zero or very small.  These results and estimates of the Hubble constant $H_o$ and deceleration parameter $q_o$ from precision redshift/magnitude data support a major assumption of this paper that the Friedmann solution to Einstein's theory of general relativity without cosmological constant is an adequate approximation to spacetime geometry and its long term evolution at quantum scales.

\end{abstract}


\section{Introduction}

The fine-structure constant $\alpha$ does not vary as Friedmann Universes evolve, a conclusion based on assessments of quantum mechanics and electrodynamics.   This research was motivated by predicted evolutionary changes in atoms and in vacuum permittivity $\varepsilon_o$. 

The spacetime geometry of the Friedmann Universe depends on the radius $a(t)$
\begin{equation}
ds^2\,	=\,c^2dt^2\,-\,a^2 (t) \left[\frac{dr^2}{\left(1-r^2\right)}\,+\,r^2\left(d\vartheta^2\,+\,sin^2\vartheta \,
		d\varphi^2\right)\right]. 
\end{equation}

Schr\"odinger (1939) found that the wavelength of every quantum wave function is proportional to $a(t)$.  This was used to explain Hubble redshift of photons from distant nebulae, one of the first clear indications of a dynamic universe.  But Schr\"odinger's result applies to every quantum system.  Atoms expand with the universe just as photons do.  Calculating the changes in atomic energy levels implied by this evolution leads directly to the constancy of $\alpha$.

$\alpha$ is the dimensionless number 
\begin{equation}
\alpha \equiv \frac {e^2}{4\pi\varepsilon_o\hbar c}\approx \frac {1}{137},
\end{equation}
where $e$ is the charge of the electron, $\varepsilon_o$ is vacuum permittivity, $\hbar$ is Planck's constant divided by $2\pi$, and $c$ is the speed of light.  

Sumner (1994) found that $\varepsilon_o$ is proportional to $a(t)$, 
\begin{equation}
\frac{\varepsilon_o (t_1)}{\varepsilon_o (t_2)}\,=\,\frac{a(t_1)}{a(t_2)}.
\end{equation}

This precisely reproduces Schr\"odinger's results for photons and atoms.  The electrodynamic wave equation gives the dependency of the speed of light on $a(t)$,
\begin{equation}
\frac{c (t_1)}{c (t_2)}\,=\,\frac{a(t_2)}{a(t_1)}.
\end{equation}
Since $e$ and $\hbar$ do not depend on $a(t)$, $\alpha$ does not depend on $a(t)$.  

Since $\alpha$ appears in all first-order perturbation formulas for atomic energy levels, comparisons of the atomic spectra of distant atoms with those in laboratories provide an experimental measure of this prediction.  For example, Bahcall et al. (2003) found  $|\alpha^{-1} d \, \alpha(t)/dt |
~<~ 2\times 10^{-13}$ yr$^{-1}$, corresponding to $ \Delta \alpha/{\alpha(0)}~=~\left(0.7 \pm 1.4\right)
\times 10^{-4}$ for quasars with $0.16 < z < 0.80$.  Other experiments also find changes that are either zero or very small within experimental errors.

The Hubble constant $H_o$ and deceleration parameter $q_o$ are determined using precision redshift/magnitude data (Riess et al. 2004).  The maximum observed redshift of 5.4 (Rhoads et al.  2004) is used to estimate the current value of the Friedmann radius $a(t_o)$ as less than $0.16$. These further support a major assumption used in this paper that the Friedmann solution to Einstein's theory of general relativity without cosmological constant is an adequate approximation to spacetime geometry and its long term evolution at quantum scales. 

\section{Quantum Wave Functions}

Schr\"odinger (1939) established that the plane-wave eigenfunctions characteristic of flat spacetime are replaced in the Friedmann universe by quantum wave functions with wavelengths directly proportional to $a(t)$.  This means that eigenfunctions change wavelength as $a(t)$ changes and the quantum systems they describe do as well.  In an expanding universe quantum systems expand.  In a contracting universe they contract.

Schr\"odinger reproduced the changes in photon and particle momenta well known from general relativity, confirming the logic of each approach and the applicability of Friedmann geometry to quantum processes.

These changes in quantum systems may equivalently be viewed as a logical consequence of the fact that the energy and momentum of an ``isolated'' system can change in general relativity when the spacetime geometry of the universe changes (Schr\"odinger 1956).  ``In an expanding space {\it all momenta decrease} \ldots for bodies acted on by no other forces than gravitation \ldots This simple law has an even simpler interpretation in wave mechanics: all wavelengths, being inversely proportional to the momenta, simply expand with space.''

Since energy is proportional to the square of momentum, quantum level energies are proportional to $a^{-2}(t)$.  The wavelengths emitted in quantum transitions are therefore proportional to $a^2(t)$ (Sumner \& Sumner 2000).  This is true for all emissions, including those from atomic states separated by the fine-structure interaction.

Bahcall et al. (2003) considered the wavelengths, $\lambda_1(t)$ and $\lambda_2(t)$, from two atomic states separated by the fine-structure interaction and proved that ``to very high accuracy \dots the difference in the wavelengths divided by the average of the wavelengths, $R$,
\begin{equation}
R(t)~=~\frac{\lambda_2(t) - \lambda_1(t)}{\lambda_1(t) + \lambda_2(t)} \, , \label{eq:defnR}
\end{equation}
is proportional to $\alpha^2$ \dots a measurement of $R$ is a measurement of $\alpha^2$ at the
epoch at which \dots lines are emitted.'' 

From Schr\"odinger's perspective each wavelength emitted in a quantum transition is proportional to $a^2(t)$.  To calculate $R(t)$ at any epoch, each $\lambda$ in equation (4) would be multiplied by an identical factor.  In every case these factors cancel and $R(t)$ is unchanged.  Since constant $R(t)$ implies constant $\alpha$, the fine-structure constant is unaffected by the coevolution of atoms, photons, and the Friedmann Universe. 

\section{Vacuum Permittivity}

By generalizing electrodynamics to the curved Friedmann spacetime of general relativity, a purely geometrical problem in 4-space, Sumner (1994) found that $\varepsilon_o$ is proportional to $a(t)$.

$\varepsilon_o(t)$ and non-relativistic quantum mechanics exactly reproduce the shifts in atomic energy levels and wavelengths of emitted photons implied by Schr\"odinger's result, i.e., the wavelengths of emitted photons are proportional to $a^2(t)$.  As the geometry evolves, the wavelengths of atomic emissions, which are proportional to $a^2(t)$, change faster than photon wavelengths, which are proportional to $a(t)$.  While both change in the same direction, i.e., both get redder or both get bluer, the difference in the magnitudes of change reverses the historical interpretation of Hubble redshift.

Hubble \& Tolman (1935) found that there is an approximately linear relationship between the magnitudes of the redshifts observed for distant galaxies and their distances from Earth. The most likely explanation  presented by Hubble \& Tolman is that galaxies are receding from the Earth. This conventional conclusion that redshifts imply recession depends critically on the assumption that the wavelengths of light emitted by atoms from distant galaxies in the past are exactly the same as the wavelength of light similar atoms emit today.   But this is not the case.  Atoms coevolve with the universe and their emissions change wavelength.  Today's standards have shifted.

If the universe were expanding, redshifted photons from distant galaxies would be measured against laboratory atoms which have redshifted more during the time the photon has taken to reach the Earth. The shift observed between the old galactic photon and the current laboratory standard would be blue, contrary to observation. Only when the universe is contracting and blue-shifted photons are compared against bluer atomic standards are the measured shifts red and consistent with Hubble's observations. Hubble redshift implies that our universe is contracting.  

A contracting Friedmann Universe must be closed, verifying Schr\"odinger's (1939) original assumption.  ``Wave mechanics imposes an a priori reason for assuming space to be closed; for then and only then are its proper modes discontinuous and provide an adequate description of the observed atomicity of matter and light.'' 

Possible changes in $\hbar$ must also be considered to determine the variation in $\alpha$.  Sumner (1994) calculated the energy changes for wave solutions to Maxwell's equations.  It was found that the energy changes were consistent with the known dependency of wavelength on $a(t)$, verifying that Planck's constant $\hbar$ relating the two remains constant as $a(t)$ changes.  Another (and particularly elegant) way of looking at the constancy of $\hbar$ is to note that $\hbar$ is the constant of proportionality between the phase of a wave function in the limiting (classical) case and the mechanical action of the physical system, a relationship that does not depend on geometry (Landau \& Lifshitz 1977). 

M\o ller (1952) also explored the dependence of $\varepsilon_o$ on spacetime geometry.  He demonstrated that the equation of continuity, implied by electrodynamics, requires that $e$ remain constant even as spacetime geometry changes.

The final term in $\alpha$ is the speed of light, $c$.  For an isotropic, homogeneous medium, the vacuum permittivity $\varepsilon_o$ equals the magnetic permeability $ \mu_o$, an equality not changed by spacetime curvature (M\o ller 1952).  The velocity of propagation, $c$, is given by
\begin{equation}
c=\left(\varepsilon_o \mu_o\right)^{-1/2}=\left(\varepsilon_o \right)^{-1}.
\end{equation}
Equation (6) and equation (3) prove the dependency of $c$ on  $a(t)$.

The variations in $\varepsilon_o$ and $c$ cancel each other and the other terms are constant.  Consequently $\alpha$ remains constant as the curvatures of Friedmann Universes evolve. 

\section{Comparisons with Experiments}

Since $\alpha$ appears in all first-order perturbation formulas for atomic energy levels, comparisons of the atomic spectra of distant atoms with those in laboratories provide an experimental measure of the theoretical prediction made here that $\alpha$ is constant.

Bahcall et al. (2003) used ``\dots the strong nebular emission lines of O~{\sc iii}, 5007~\AA\ and 4959~\AA, to set a robust upper limit on the time dependence of the fine structure constant. We find $|\alpha^{-1} d \, \alpha(t)/dt |~<~ 2\times 10^{-13}$ yr$^{-1}$, corresponding to $ \Delta \alpha/{\alpha(0)}~=~\left(0.7 \pm 1.4\right)\times 10^{-4}$ for quasars with $0.16 < z < 0.80$ \dots ''  This research was a continuation of work done by Bahcall \& Salpeter (1965) significantly improving the accuracy, but arriving at the same conclusion.

Bahcall et al. (2003) contains a useful review of methods used to measure changes in $\alpha$.  While most conclude that there is no measurable change in $\alpha$, some conclude that there is statistically significant change.  For example,  Murphy et al. (2003) find,  ``$\Delta\alpha/\alpha = (-0.543 \pm 0.116) \times 10^{-5}$, representing 4.7\,$\sigma$ evidence for a varying $\alpha$.''   On the other hand, Srianand et al. (2004) conclude that their analysis  ``\dots gives a 3$\sigma$ limit, $-2.5\times 10^{-16} ~{\rm yr}^{-1}\le(\Delta\alpha/\alpha\Delta t) \le+1.2\times 10^{-16}~{\rm yr}^{-1}$, for the time variation of $\alpha$, that forms the strongest constraint obtained based on high redshift quasar absorption line systems.'' 

If $\alpha$ does in fact vary in time (or in space), the simplifying assumption that spacetime is adequately modeled by Friedmann geomtery would need to be closely reexamined.  

\section{Discussion}

Two independent lines of theoretical reasoning are presented that conclude the fine-structure constant does not evolve with Friedmann geometry.  Key assumptions made by Schr\"odinger and Sumner are: 
\begin{enumerate}

\item The Friedmann solution to Einstein's theory of general relativity without cosmological constant is an adequate approximation to spacetime geometry and its long term evolution at quantum scales (Schr\"odinger and Sumner).

\item Classical electrodynamics may be generalized to the curved spacetime of general relativity (Sumner).

\item Non-relativistic quantum mechanics is an adequate model for atoms, with fine-structure splitting calculated in the usual way (Sumner). 

\item Relativistic quantum mechanics can be used to determine the relationship between quantum wave functions and Friedmann geometry (Schr\"odinger).

\end{enumerate}

It has been shown that these assumptions require that atoms coevolve with spacetime geometry.  While this conclusion does not agree with Einstein's original assumption that physical ``rigid rods'' measure mathematical space, it is not in conflict with general relativity per se as Einstein (1949) emphasized.

\begin{quote}Can a spectral line be considered as a measure of a ``proper time'' ($EigenZeit$) $ds$ ($ds^2 = g_{ik}dx_idx_k$), (if one takes into consideration regions of cosmic dimensions)?  Is there such a thing as a natural object which incorporates the ``natural-measuring-stick'' independently of its position in four-dimensional space? The affirmation of this question made the invention of the general theory of relativity \begin{it}psychologically\end{it} possible; however this supposition is logically not necessary.  For the construction of the present theory of relativity the following is essential:

(1)  Physical things are described by continuous functions, field-variables of four co-ordinates.  As long as the topological connection is preserved, these latter can be freely chosen.

(2)  The field-variables are tensor-components; among the tensors is a symmetrical tensor $g_{ik}$ for the description of the gravitational field.

(3)  There are physical objects, which (in the macroscopic field) measure the invariant $ds$.

If (1) and (2) are accepted, (3) is plausible, but not necessary.  The construction of mathematical theory rests exclusively upon (1) and (2). A $complete$ theory of physics as a totality, in accordance with (1) and (2) does not yet exist.  If it did exist, there would be no room for the supposition (3). For the objects used as tools for measurement do not lead an independent existence alongside of the objects implicated by the field-equations.\end{quote}

There is  ``no room for the supposition (3)''.  Rigid rods do not have an existence independent of spacetime geometry.  Atoms (and meter sticks comprised of them) coevolve with the Friedmann Universe. 

Spacetime curvature and its rate of change can be estimated using redshift/magnitude measurements of supernovae.  The ``gold'' selection of redshifts of 156 supernovae from Riess et al. (2004) has been analyzed.  Using the mathematical modifications required by contraction (Sumner \& Vityaev  2000, Sumner 2004) it was found that $H_o\,=\,-66.6\, km\,s^{-1}\,Mpc^{-1}$ and $q_o=0.5+|\epsilon|$, where $\epsilon \approx 0$ (abbreviated as $q_o\,\gtrsim \,0.5$).  The standard error for this fit is $0.287$, compared with $0.245$, the average of the stated data errors.  Figure 1 illustrates this fit and Figure 2 shows the fit sensitivity to parameter choice.  Applying the same procedure to wider data sets from Riess et al. (2004) gives essentially the same result but with higher errors. 

\begin{center}
\psfig{figure=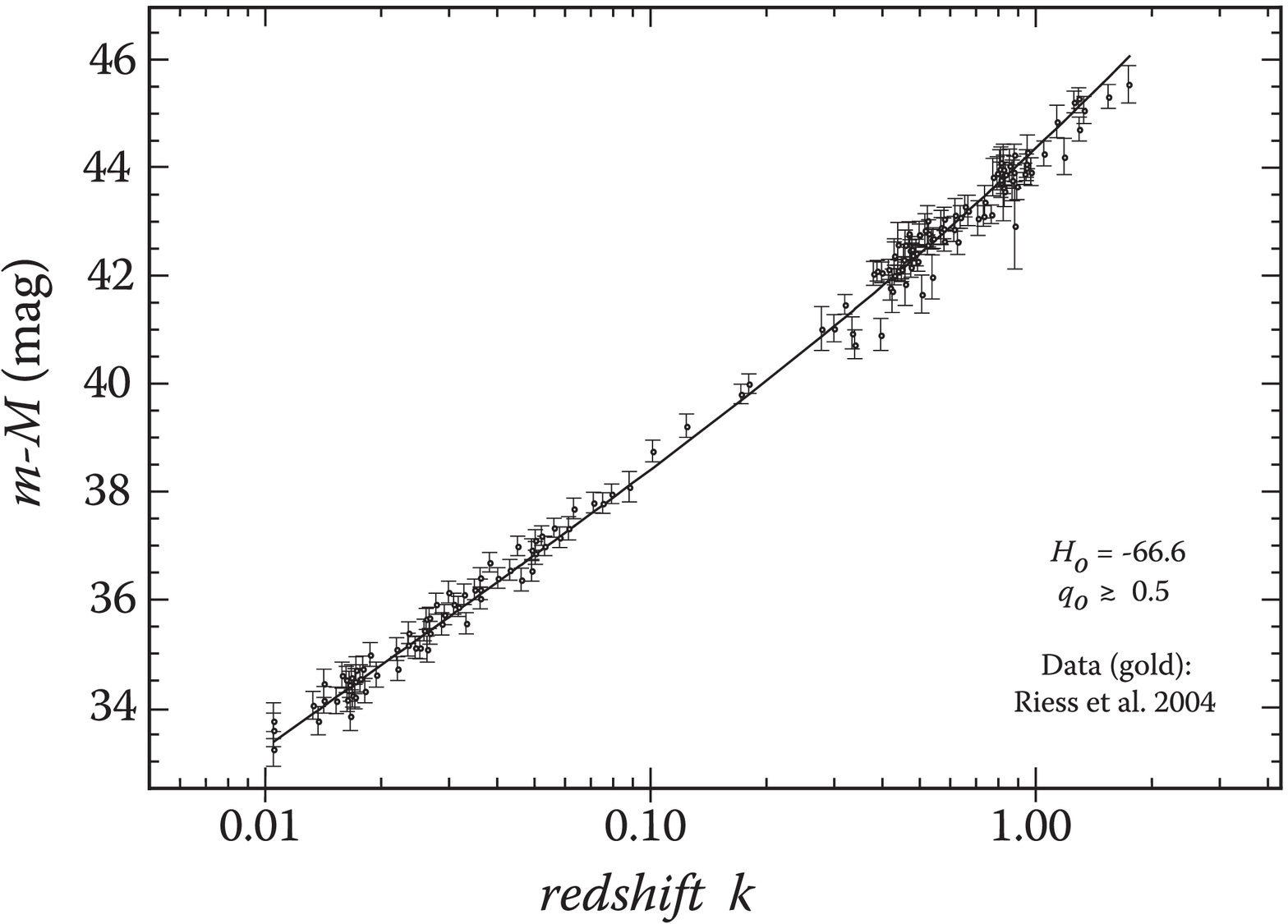,width=12.5 cm}
\figcaption[fig1]{Redshifts and magnitudes for 156 supernovae (the ``gold'' data of Riess et al. 2004) and a fit with the parameters $H_o\,=\,-66.6\, km\,s^{-1}\,Mpc^{-1}$ and $q_o\,\gtrsim \,0.5$}
\end{center}

\begin{center}
\psfig{figure=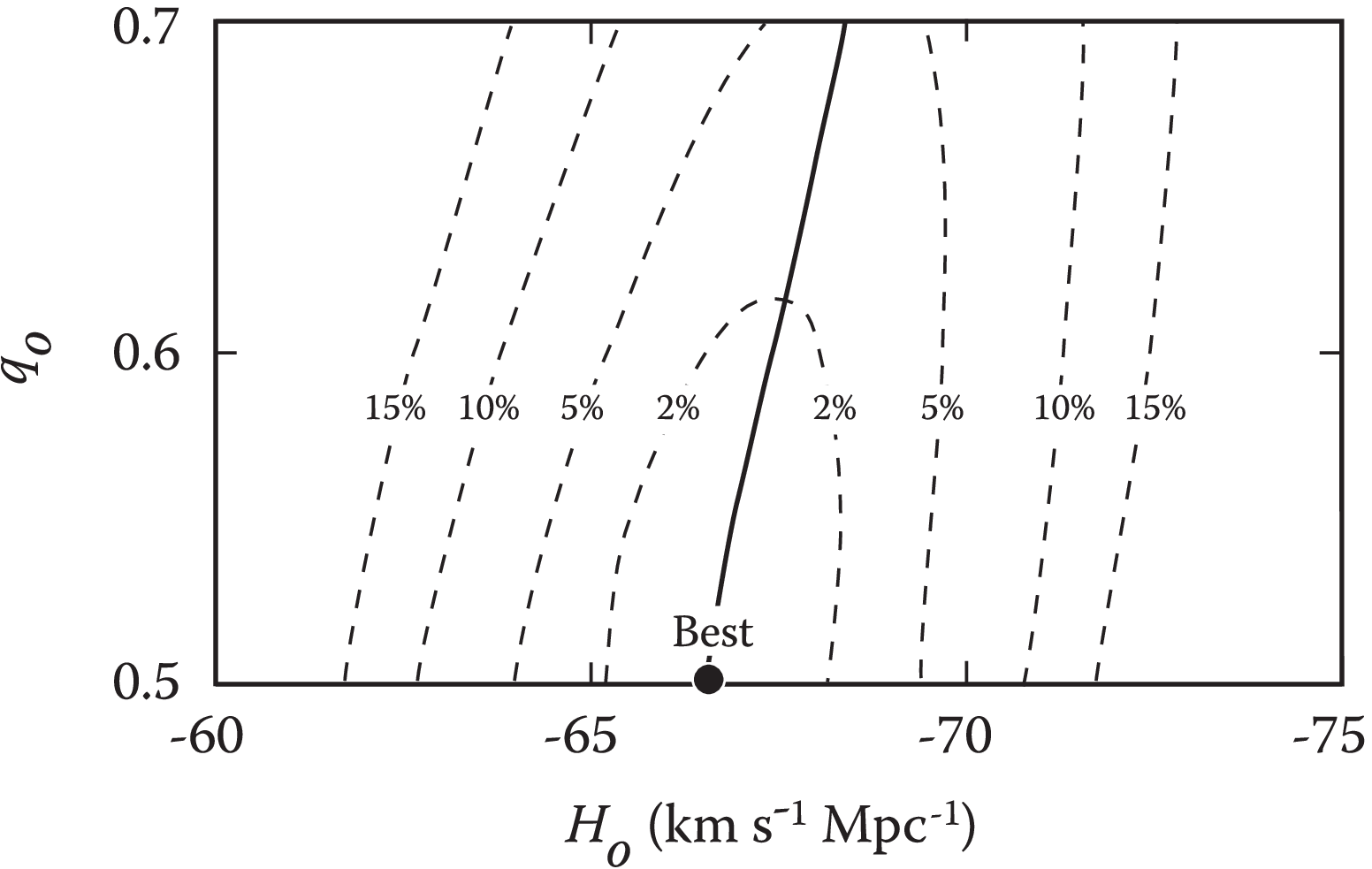,width=8.0 cm}
\figcaption[fig2]{Parameters within 2, 5, 10, and 15\% of the best fit  $H_o\,=\,-66.6\, km\,s^{-1}\,Mpc^{-1}$ and $q_o\,\gtrsim \,0.5$ for the ``gold'' data of Riess et al. 2004}
\end{center}

For $H_o =-66.6\, km\,s^{-1}\,Mpc^{-1}$ and $q_o\, \gtrsim \, 0.5$,  the time remaining is about 15 billion years.  Since the minimum standard error is when $q_o\, \gtrsim \, 0.5$, only a minimum limit for age for the universe may be estimated.   By using the maximum observed redshift of 5.4 (Rhoads et al. 2004), $q_o$ must be $0.59$ or less which implies that $a(t_o) <  0.16$ and the universe is at least 700 billion years old (Sumner 2004).  Figure 3 illustrates the evolution of a Friedmann universe with these parameters.

\begin{center}
\psfig{figure=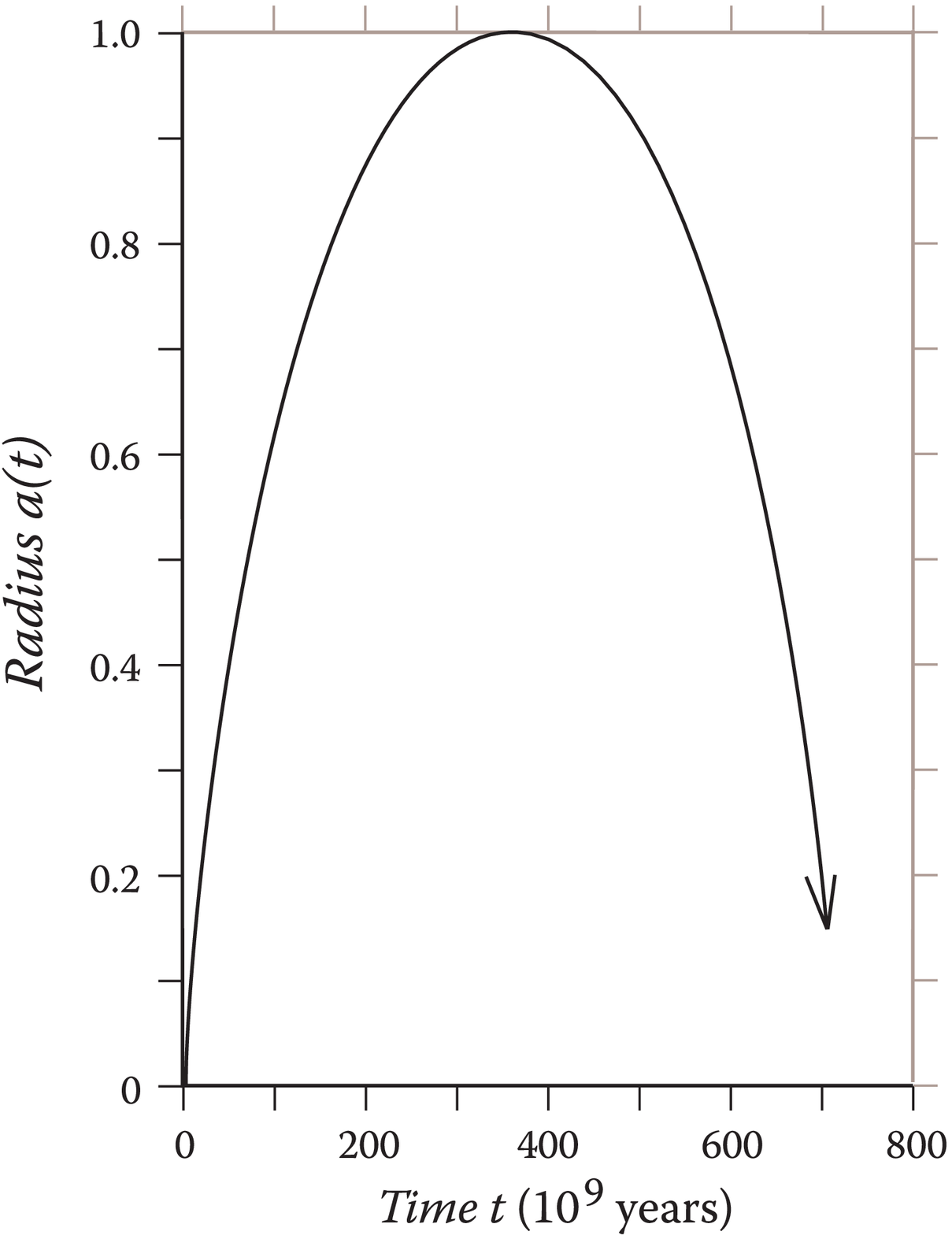,width=6.0 cm}
\figcaption[fig3]{Friedmann Universe for $H_o\,=\,-66.6\, km\,s^{-1}\,Mpc^{-1}$ and $q_o\,=\,0.59$ with the arrow tip at the current epoch}
\end{center}

Only $H_o$ and $q_o$ were varied to fit this redshift/magnitude data and no cosmological constant was included.  The ``acceleration'' of the universe this data reveals is real and reflects the acceleration of contraction as the universe collapses.   While this redshift data does not preclude the use of a cosmological constant, it does not present a compelling reason for its inclusion.  

\section{Conclusions}

Relativistic quantum mechanics was used to show that the fine-structure constant $\alpha$ remains constant and that atoms and photons coevolve with Friedmann spacetime geometry.   These conclusions follow directly from Schr\"odinger's insight that every quantum wave function is proportional to the radius of Friedmann spacetime, a result beautifully consistent with the original observations made by de Broglie (1929). 

\begin{quote}We thus find that in order to describe the properties of Matter, as well as those of Light, we must employ waves and corpuscles simultaneously.  We can no longer imagine the electron as being just a minute corpuscle of electricity:  we must associate a wave with it. And this wave is not just a fiction:  its length can be measured and its interferences calculated in advance.\end{quote}

The dependence of vacuum permittivity $\varepsilon_o$ on spacetime curvature implies both a constant $\alpha$ and the same coevolution required by relativistic quantum mechanics.

Since $\alpha$ appears in all first-order perturbation formulas for atomic energy levels, comparisons of the atomic spectra of distant atoms with those in laboratories provide an experimental measure of this prediction.  Most  experiments find changes in $\alpha$ that are either statistically zero or very small.

Friedmann spacetime is also shown to be consistent with precision redshift/magnitude data (Riess et al. 2004).  The assumption made by both Schr\"odinger and Sumner that \textit{the Friedmann solution to Einstein's theory of general relativity without cosmological constant is an adequate approximation to spacetime geometry and its long term evolution at quantum scales} is well-founded.

\bigskip 
\bigskip 
\bigskip 

The author thanks D.Y. Sumner for many discerning insights and helpful suggestions.

\end{document}